\documentclass[12pt]{iopart}

\usepackage{graphicx}
\usepackage{subfig}
\usepackage{multirow}
\begin{document}

\title{Hydrodynamic flow in the vicinity of a nanopore induced by an applied voltage}
\author{Mao Mao$^1$, Sandip Ghosal$^2$ and Guohui Hu$^3$}

\address{$^1$ Department of Mechanical Engineering, Northwestern
University, 2145 Sheridan Road, Evanston, IL, 60208, USA}

\address{$^2$ Departments of Mechanical Engineering and Engineering Sciences 
\& Applied Mathematics,  Northwestern University, 2145 Sheridan Road, Evanston, IL, 60208, USA}

\address{$^3$ Shanghai Institute of Applied Mathematics and
Mechanics, 
Shanghai University, 149 Yanchang Road, Shanghai 200072, P. R. China}

\ead{MaoMao2015@u.northwestern.edu}

\begin{abstract}
Continuum simulation is employed to study ion transport and fluid flow through a nanopore in a solid-state membrane under an applied potential drop. Results show the existence of concentration polarization layers on the surfaces of the membrane. The nonuniformity of the ionic distribution gives rise to an electric pressure that drives vortical motion in the fluid. There is also a net hydrodynamic flow through the nanopore due to an asymmetry induced by the membrane surface charge. The qualitative behavior is similar to that observed in a previous study using molecular dynamic simulations. The current--voltage characteristics show some nonlinear features but are not greatly affected by the hydrodynamic flow in the parameter regime studied.  In the limit of thin Debye layers, the electric  resistance of the system can be characterized using an equivalent circuit with lumped parameters. Generation of vorticity can be understood qualitatively from elementary considerations of the Maxwell stresses. However, the flow strength is a strongly nonlinear function of the applied field. Combination of electrophoretic and hydrodynamic effects can lead to ion selectivity in terms of valences and this could have some practical applications in separations.
\end{abstract}

\section{Introduction}
Ionic conduction through nanometer-sized channels or pores is a common theme in biological systems as well as in various manufactured materials such as membranes and 
synthetic nanopores
\cite{doyle1998thestructure,rhee2006nanopore,venkatesan2011nanopore,branton2008thepotential,roux2004theoretical,demming2012nanoporestextemdashthe,Liu2009,Qiao2009,Lu2012,Kong2006}. 
Typical synthetic nanopore systems have a length scale of a few nanometers to tens of nanometers. This is still sufficiently large compared to molecular sizes that continuum simulations remain a powerful tool for studying them~\cite{Ai2011,Ai2011a,Ai2010,Stein2010,Lee2012,Liu2007}, even though, its accuracy decreases progressively as the pore size approaches that of biological pores. This is because molecular dynamic (MD) simulations at such mesoscales are computationally very expensive.

The objective of this paper is to study the hydrodynamic flow that arises in the vicinity of a single cylindrical nanopore when a voltage is applied across the membrane. The role of electrically induced hydrodynamic  flows near pores, arrays of pores or semipermeable membranes have been addressed by many authors~\cite{Chang2011}. Hydrodynamics is also implicated in 
polymer translocation through nanopores where viscous resistance is shown to determine translocation times \cite{ghosal2006electrophoresis,ghosal2007effect,Ghosal2007} and stall forces \cite{VanDorp2009}
applied by optical traps to immobilize DNA in nanopores against the electric driving force.

Under the appropriate set of circumstances, hydrodynamics at the entrance or the exit of a nanopore has a strong influence on ion transport~\cite{Chang2011,Yossifon2010,Yossifon2008,Chang2009}.  
Nanoslots and nanochannels have been shown to exhibit 
ion selectivity similar to that of a perm--selective membrane. Current--voltage curves for such systems show a nonlinear saturation of the current at high voltages followed by an ``overlimiting'' regime dominated by convective transport by fluid vortices~\cite{Chang2009}. In the case of perm selective membranes, or equivalently for a dense array of nanochannels in a membrane, the current saturation is caused by the loss of carrier ions in the concentration polarization layer (CPL) adjacent to the membrane 
\cite{Levich,Rubinstein1979,Ben2002,Yariv2009}. The hydrodynamic flow has been attributed to a loss of stability in the CPL 
at higher voltages \cite{Rubinstein2000,ZALTZMAN2007} that generates a vortex array and selects a diffusion length scale  that is different from the one in the quiescent Ohmic regime~\cite{Yossifon2009}. The overlimiting current is a signature of this instability. These postulated micro-scale vortices have subsequently been confirmed through direct observation~\cite{Yossifon2008,Rubinstein2008}. A related electrokinetic effect of relevance is what has been termed  `electro--osmosis of the second kind' by Dukhin \cite{Dukhin1991}.  Here the flow is driven by the field component tangential to a curved surface whereas the 
space charge layer (SCL) itself is created by the normal component of the field. In the case of an isolated nanopore with strongly overlapping Debye layers this results in corner vortices due to a field--focusing effect and the limiting current plateau in the current voltage characteristic is eliminated~\cite{Yossifon2010_1}. Such corner vortices are also expected at corners in micro-channels due to leakage of the electric field into the dielectric \cite{Yossifon2006added}.

In our previous paper \cite{Hu2012} we utilized molecular dynamics (MD) simulation to study the transport and flow field in a graphene 
sheet nanopore. We observed concentration polarization, corner vortices and non--linear current--voltage relation for the graphene sheet nanopore system. The results are in qualitative agreement with physical expectations and theory. In this paper, we utilize continuum level simulation for transport and fluid flow in a similar nanopore system. Such an approach has the advantage of computational speed but certain features, for example steric effects, cannot be captured. We are interested in the overlap between the continuum and MD approaches which we study by comparing our simulations to the previous MD results. The rest of the paper is organized in this way: in section 2, we present our modeling and describe our numerical approach. The results are presented in section 3. Analysis and discussions are in section 4.

\section{Method}
\subsection{Problem formulation}
\begin{figure}[h]
\centering
\includegraphics[width=0.8\textwidth]{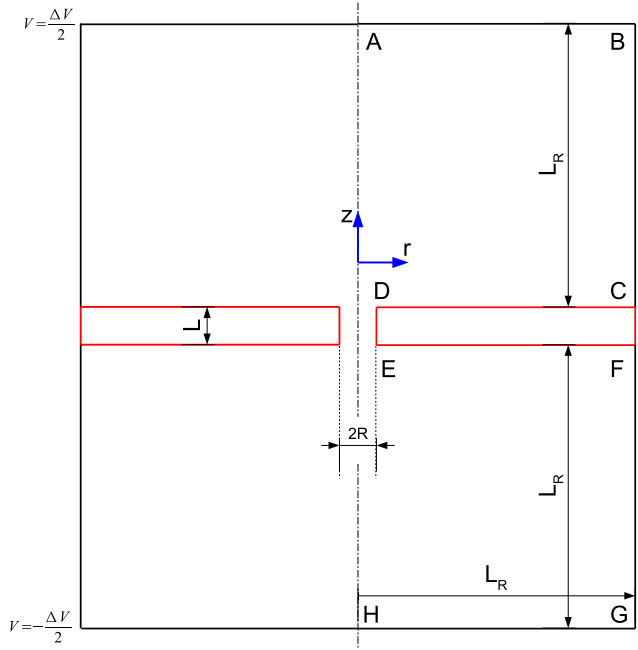}
\caption{A sketch of the nanopore system showing cylindrical coordinates. The system is axisymmetric.}
\label{fig:system}
\end{figure}

Our nanopore system consists of a cylindrical pore of radius $R$ drilled on a membrane of thickness $L$. A cylindrical coordinate system is used. The pore connects two reservoirs that have equal radius and depth, $L_R$. A sketch of the nanopore system is shown in Figure \ref{fig:system}. The membrane surface is indicated by the lines: CD, DE and EF. 

The bath is taken as a KCl solution where the salt is fully dissociated. The concentration of either ionic species is taken as $c_0$. 
We will assume, for simplicity, that the membrane has a fixed surface charge of density $\Sigma$ and ignore any 
changes induced by shifts of the ionization equilibrium (from various effects such as changes in salt concentration and pH).
For the purpose of estimation, a rule of thumb is that $\Sigma$ cannot exceed the value of one electron charge per Bjerrum length \cite{Chang2011}. Due to attraction of counter--ions, in this case K$^+$, an electric double layer (EDL) forms along the charged surfaces (CD, DE and EF). An external electric field is used to generate current through the nanopore. The field will induce its own space charge along surfaces due to concentration polarization (CP) thereby modifying the existing EDL. The action of the tangential electric field on the space charges drags the fluid along the membrane resulting in electroosmotic flow (or EOF). The Nernst-Planck-Poisson (NPP)--Stokes system of equations describes the coupling between ion transport, electric field and hydrodynamics.

Assuming that a stationary state has been reached, the NPP--Stokes system of equations in the solution domain consists of the continuity equation:
\begin{equation}
\nabla \cdot \mathbf{u}=0
\label{eq:incompressible}
\end{equation}
the Stokes equation:
\begin{equation}
-\nabla p + \mu \nabla^2 \mathbf{u} - e (z_1 c_1+z_2 c_2) \nabla V=0
\label{eq:Stokes}
\end{equation}
the Nernst--Planck equations for cations and anions:
\begin{equation}
\mathbf{N}_i =  - D_i \nabla c_i - z_i \frac{D_i}{k_BT} e c_i \nabla V+\mathbf{u} c_i, \qquad i=1,2
\label{eq:NP}
\end{equation}
\begin{equation}
\nabla \cdot \mathbf{N}_i = 0, \qquad i=1,2
\end{equation}
and Poisson equation for the electric potential:
\begin{equation}
-\epsilon \nabla^2 V = e (z_1c_1+z_2c_2).
\label{eq:Poisson}
\end{equation}
The fluid flow is governed by the Stokes equation with an additional electrostatic volume force term, since the inertia term in the 
Navier--Stokes equation is inconsequential at the nanoscale. 
Here $\mathbf{u}=u \hat{e}_r + w \hat{e}_z$ is the flow velocity, $\hat{e}_r$ and $\hat{e}_z$ being the unit vectors in the radial ($r$) and 
axial ($z$) directions respectively, $p$ is the fluid pressure and $c_i$ ($i=1,2$) is the concentration of cation (K$^+$) and anion (Cl$^-$). The electric potential is denoted by $V$ and $\rho$, $\mu$, $\epsilon$ are the density, dynamic viscosity and permittivity of the solution respectively.   The diffusivity of the $i$th species is $D_i$ and $z_i$ is the valence ($z_1=1$ and $z_2=-1$ for KCl). The flux of species $i$ is denoted by $\mathbf{N}_{i}$ and consists of contributions from diffusion, electrophoresis and convection. The ion mobility ($u_i$) in the Nernst--Planck model obeys the Einstein relation \cite{Einstein1905, Einstein1905_2} $u_i=D_i / (k_B T)$ where 
$k_B$ is the Boltzmann constant and $T$ is the absolute temperature, taken to be 300 $K$.
In the membrane domain, we solve the Laplace equation,
\begin{equation}
\epsilon_s \nabla^2 V = 0
\end{equation}
where $\epsilon_s$ is the permittivity of the membrane material.

The boundary conditions for the system are as follows: assuming the reservoirs have a much larger volume than the pore, at BC and FG, the concentration gradient ${\partial c_i}/{\partial r}$, the radial velocity $u$ and the radial electric field $-{\partial V}/{\partial r}$ are set to zero. At AB and GH, $V=\pm {\Delta V}/{2}$ respectively and $c_i$ is set equal to the bulk concentration $c_0$ while fluid pressure $p$ is set to a 
constant value ($P_0$). At the membrane surfaces CD, DE and EF a no--penetration condition is used for ions while the no--slip condition applies for the flow. Applying the Gauss theorem in an infinitely thin control volume across the surface, we get the jump condition for normal component of the  electric field ($E_n$) which is related to the surface charge density ($\Sigma$) as $[\epsilon E_{n}] = \Sigma$. The tangential component of the field ($E_t$) is continuous, 
$[ E_{t} ] = 0$. Here the brackets $[ \quad ]$ indicate the jump in the quantity enclosed across the membrane.

We use the finite element package COMSOL MULTIPHYSICS to solve the NPP--Stokes system numerically. The triangular mesh is selected 
with refinement near the membrane surface CD, DE and EF to ensure that the Debye layer is well resolved. The refinement parameter is set such that the maximum mesh size in this region never exceeds one sixth the Debye length. The results have been checked to ensure mesh independence. Small local fillets  are used at D and E to remove the sharp corners.
In a  related paper by Yossifon {\it et al} \cite{Yossifon2010} finite element simulations are used to study a 
set up very similar to ours. However, they decouple the electrostatics and ion transport problems from the fluid flow by neglecting 
the convective flux in their ion transport equations. We do not introduce such an approximation.

\subsection{Dimensionless form}
Results in this paper are reported in terms of dimensional quantities in order to facilitate correspondence with the 
experimental literature. Nevertheless, we present the dimensionless 
form of the equations here in order to identify the principal dimensionless parameters that govern the physical behavior.
We use the bulk concentration $c_0$ as the scale for $c_i$. Electric potential is scaled with $k_BT/e$. The reference 
length scale is chosen as the pore radius, $R$. Normalizing the velocity, $\mathbf{u}$, with $U_0=D_1/R$ and pressure, $p$, with $\mu U_0/R$, we get the dimensionless form of the governing equations:

\begin{equation}
\nabla^* \cdot \mathbf{u}^* =0
\end{equation}
\begin{equation}
-\nabla^* p^* +\nabla^{*2} \mathbf{u}^* - \xi (z_1 c_1^*+z_2 c_2^*)\nabla^* V^* =0
\end{equation}
\begin{equation}
\nabla^* \cdot \left(-\nabla^* c_i^* - z_ic_i^*\nabla^* V^* + \frac{D_1}{D_i} \mathbf{u}^*c_i^* \right)=0, \qquad i=1,2
\end{equation}
\begin{equation}
-\eta^2 \nabla^{*2} V^* = \frac{z_1c_1^*+z_2c_2^*}{2}
\label{eq:normPoisson}
\end{equation}
where $\eta={\lambda_D}/{R}$ and $\xi=(c_0k_BT R^2)/(D_1 \mu)$, where $\lambda_D$, the Debye length, is defined as 
$\lambda_D=\sqrt{(\epsilon k_BT)/(2e^2 c_0)}$. All the dimensionless variables and derivatives are superscripted with $*$. 

Two dimensionless parameters emerge. The first, $\eta={\lambda_D}/{R}$ is the ratio of Debye length to the length scale $R$. It is an indication of how thin the Debye layer is in comparison with the geometric scale.  If $\eta$ is small, meaning that the Debye layer is thin relative to the geometric length scale, the right hand side of the Poisson equation (\ref{eq:normPoisson}) will be close to zero implying that the electrolyte is close to being electroneutral. Conversely, if $\eta \sim 1$, as is expected in most nanofluidic systems, Debye layers would overlap within the nanopore. The second dimensionless parameter is 
\begin{equation}
\xi=\frac{c_0k_B T R^2}{D_1 \mu}=\frac{c_0 R^2}{\mu u_{m,1}}
\end{equation}
where $u_{m,1}$ is the mobility of cations.
Multiplying by the factor $eE_0$ in both the numerator and denominator ($E_0$ representing the strength of the applied electric field)
\begin{equation}
\xi=\frac{c_0 e R^2E_0/\mu}{u_{m,1}eE_0}.
\end{equation}
Here $u_{m,1}e E_0$ is the velocity a K$^+$ ion acquires under $E_0$, or the electrophoretic velocity. The quantity $c_0eR^2E_0/\mu$ also has a dimension of velocity and can be interpreted as the velocity the fluid acquires under $E_0$ because of the space charge density it carries, or an electroosmotic velocity. 
Thus, $\xi$ is a ratio between the two velocities representing electrophoretic and electroosmotic effects. It could also be related to the electric Reynolds number $R_{E}=\xi^{-1}$, defined as the ratio of the time scale of charge convection by flow to a charge relaxation time set by Ohmic conduction \cite{Feng08061999}. 
For $R=2.5$ nm and $c_0=0.1$ M, $\eta\approx 0.4$ and 
$\xi \approx 0.9$, indicating that both finiteness of the Debye length and convective transport needs to be taken into account.

\section{Results}
The results of the simulation depend on the geometry of the system ($R$, $L$ and $L_R$), the bulk salt concentration $c_0$, the membrane surface charge density $\Sigma$ and applied voltage bias $\Delta V$. In this paper we fix the geometry of the nanopore system. We simulate a system corresponding to $R=2.5$ nm and $L=5$ nm. $L_R$ is set to be $10L$ to make sure it is large enough compared to the pore.  
 The bulk concentration $c_0$, surface charge density $\Sigma$ and external voltage $\Delta V$ are varied. Other physical properties are as follows. The relative permittivity of fluid and membrane domains are respectively 80 (for water) and 3.9 (for silica). Ion diffusivity $D_1=1.95\times10^{-9}$ m$^2/$s while $D_2=2.03\times10^{-9}$ m$^2/$s. Table \ref{table:parameters} lists all the parameters used in our simulations.
\begin{table}
\centering
\begin{tabular}{|c|c|c|c|c|c|c|}
\hline 
\multicolumn{2}{|c|}{Geometry} & \multicolumn{3}{c|}{Physical} & Relative & Diffusivity \\ 
\cline{1-5}
$R$(nm) & $L$(nm) & $c_0$($M$) & $\Delta V$($V$) & $\Sigma$($C/m^2$) & permittivity &  (10$^{-9}$ m$^2/$s)\\ 
\hline 
\multirow{4}{*}{2.5} & \multirow{4}{*}{5} & 0.001 & -2.0 & -0.01& 80(water) & 1.95(K$^+$)  \\ 
\cline{3-3} \cline{5-7}
 & & 0.01 & \multirow{2} {*}{to} & 0 & 3.9(silica) & 2.03(Cl$^-$)  \\
\cline{3-3} \cline{5-7}
 & & 0.1 & & 0.01 & &  \\
\cline{3-3} \cline{5-7}
 & & 1.0 & 2.0 & & &  \\
\hline
\end{tabular} 
\caption{Parameters used in the simulations}
\label{table:parameters}
\end{table}

\subsection{Concentration, electric field and flow field}

\begin{figure}[h]
\centering
\subfloat[The steady state concentration distribution (normalized by the bath concentration, $c_0$) in the nanopore system. The system is axisymmetric. For convenience, the left panel shows the concentration distribution of potassium ions, K$^+$ and the right panel is used for chloride ions Cl$^-$.]{
				\includegraphics[width=0.75\textwidth]{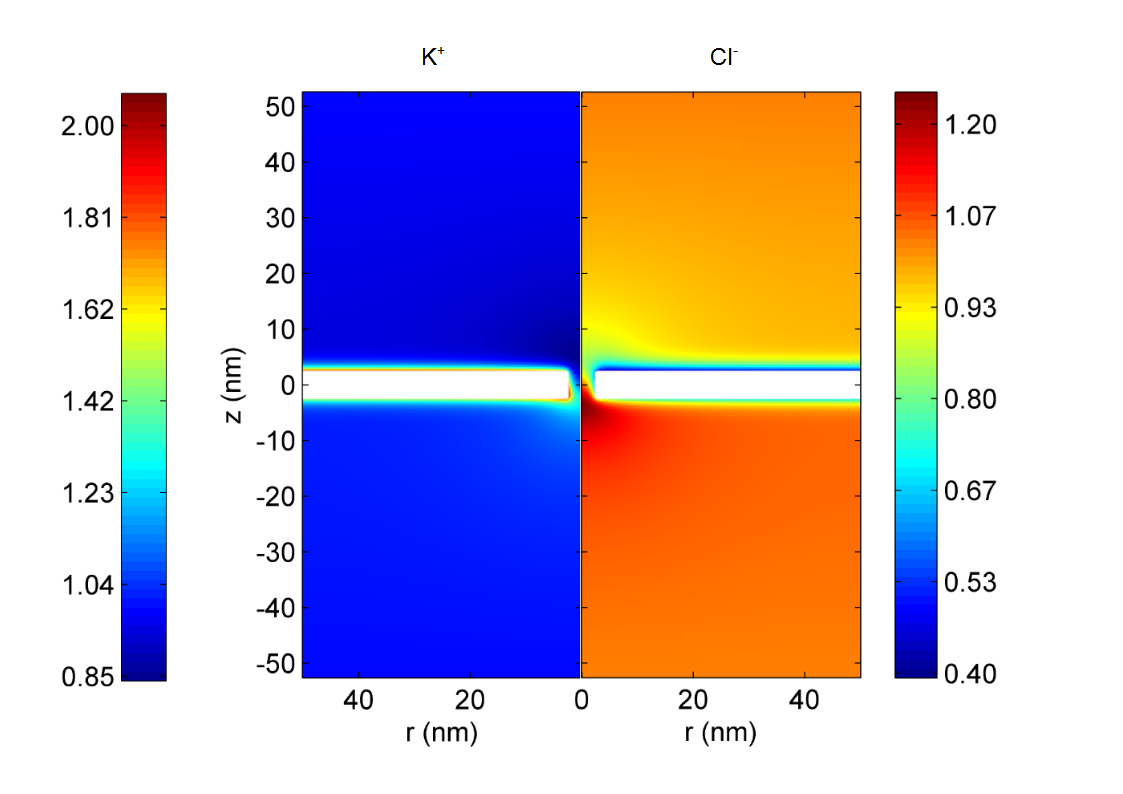}}
\\
\subfloat[Normalized ion concentrations along line $r=0$ (pore center line)]{
				\includegraphics[width=0.45\textwidth]{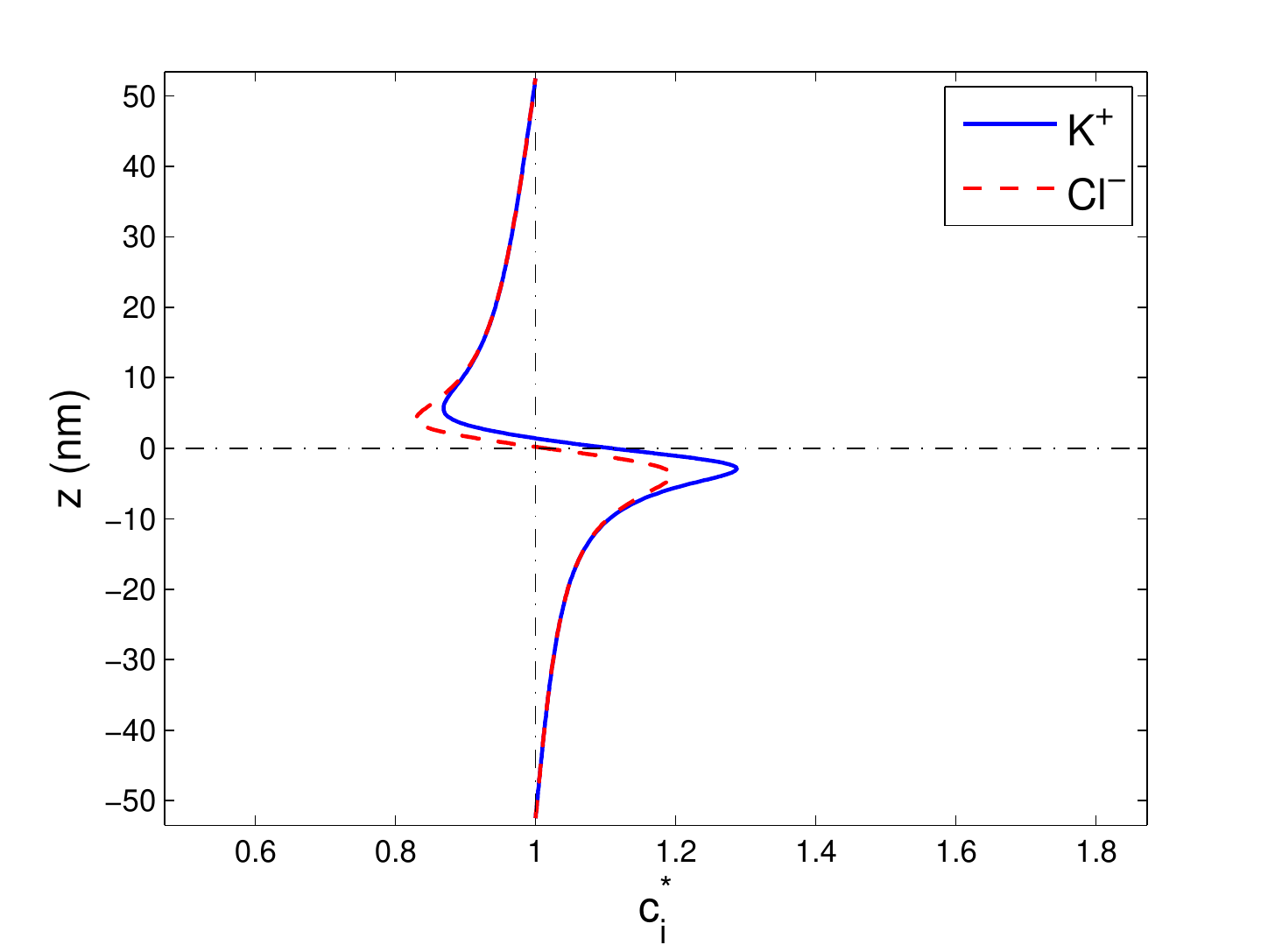}}\quad
\subfloat[Normalized ion concentrations along line $r=30$ nm (faraway from pore)]{
				\includegraphics[width=0.45\textwidth]{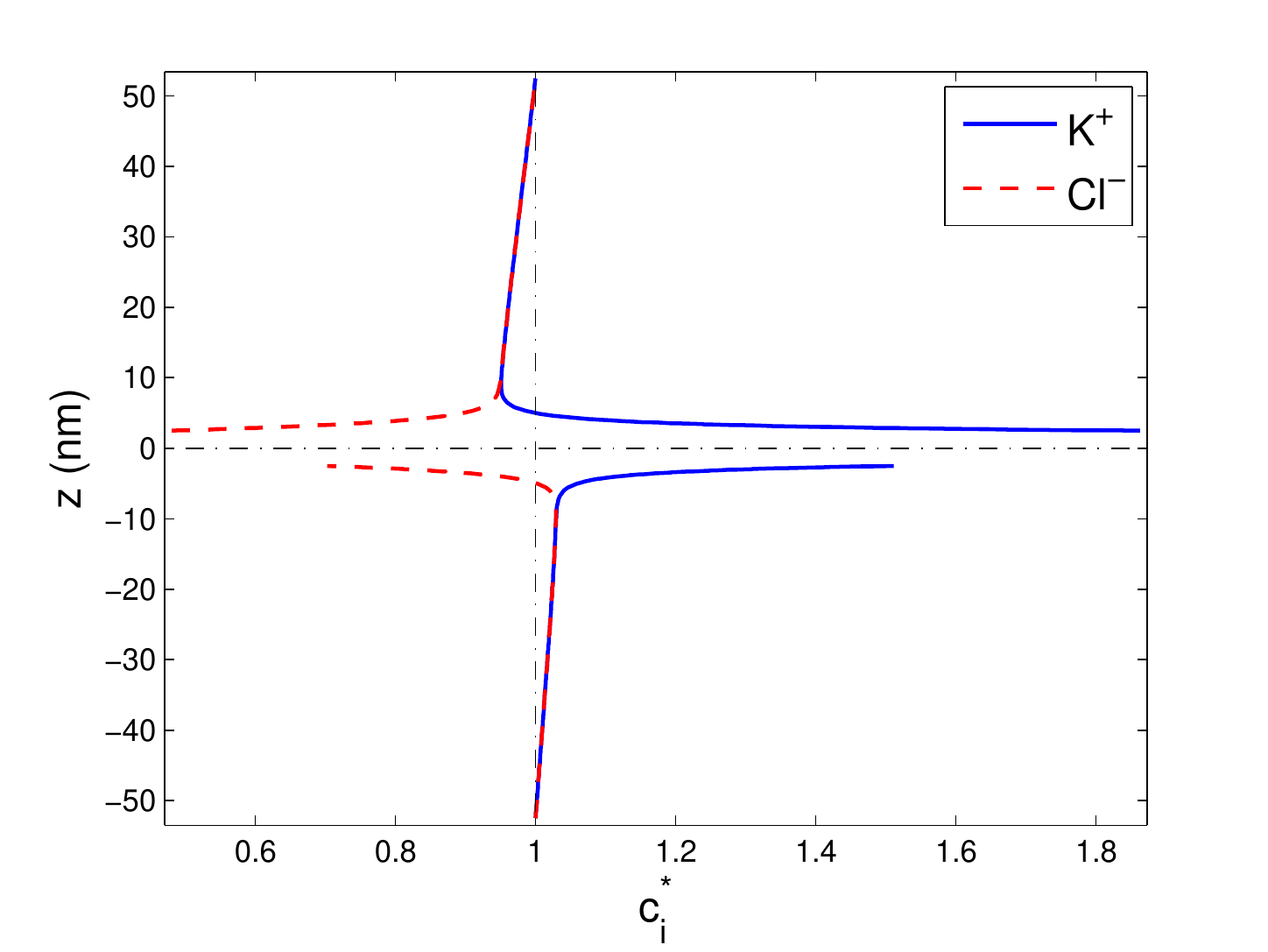}}
        \caption{Equilibrium ionic distributions in the nanopore system. The parameters are: $\Sigma=$-0.01 C/m$^2$ $\Delta V=$0.4 $V$ and $c_0=$0.1 $M$.}
\label{fig:concentration}
\end{figure}

\begin{figure}
\centering
\begin{tabular}{cc}
\multicolumn{2}{c}{
\subfloat[The electric potential distribution. The numerical values in the color map are in Volt and field lines are 
                shown in white.]{
                \includegraphics[width=0.75\textwidth]{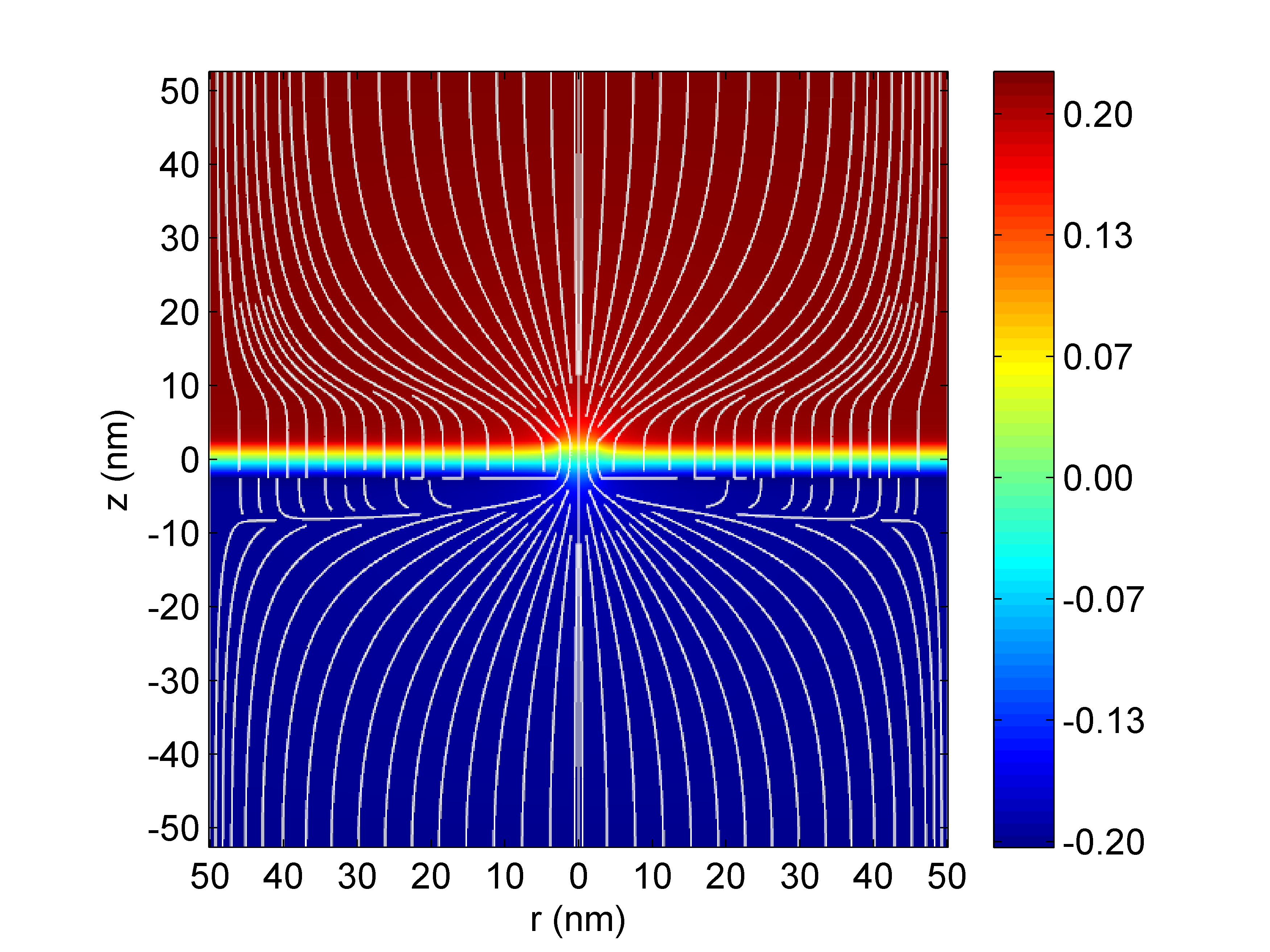}}} \\ 
\subfloat[Distribution of the electric potential along the z--axis.]{
				\includegraphics[width=0.45\textwidth]{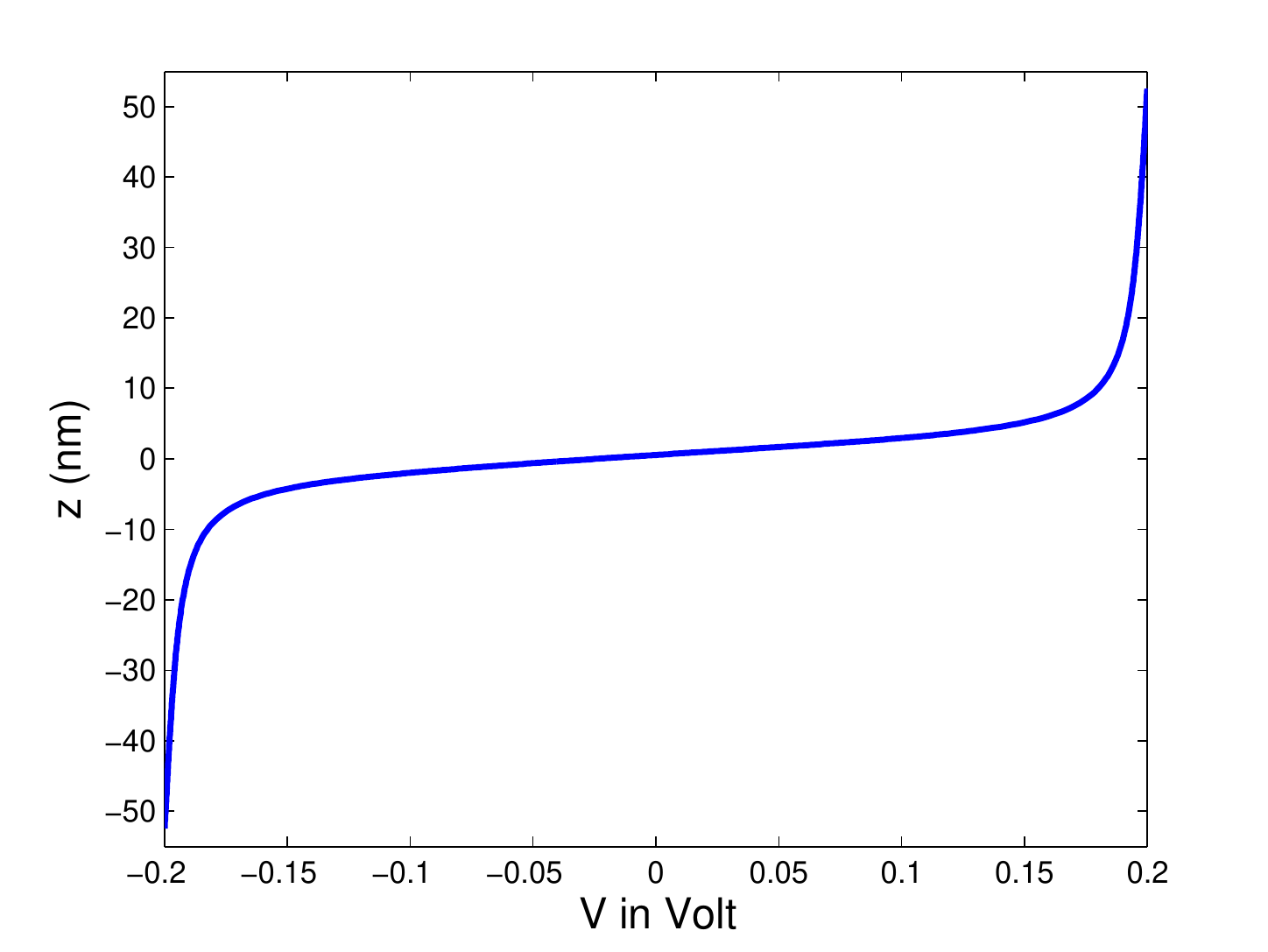}} & 
\subfloat[The radial distribution of the electric potential along the upper and lower membrane surface ($z=\pm {L}/{2}$).]{
				\includegraphics[width=0.45\textwidth]{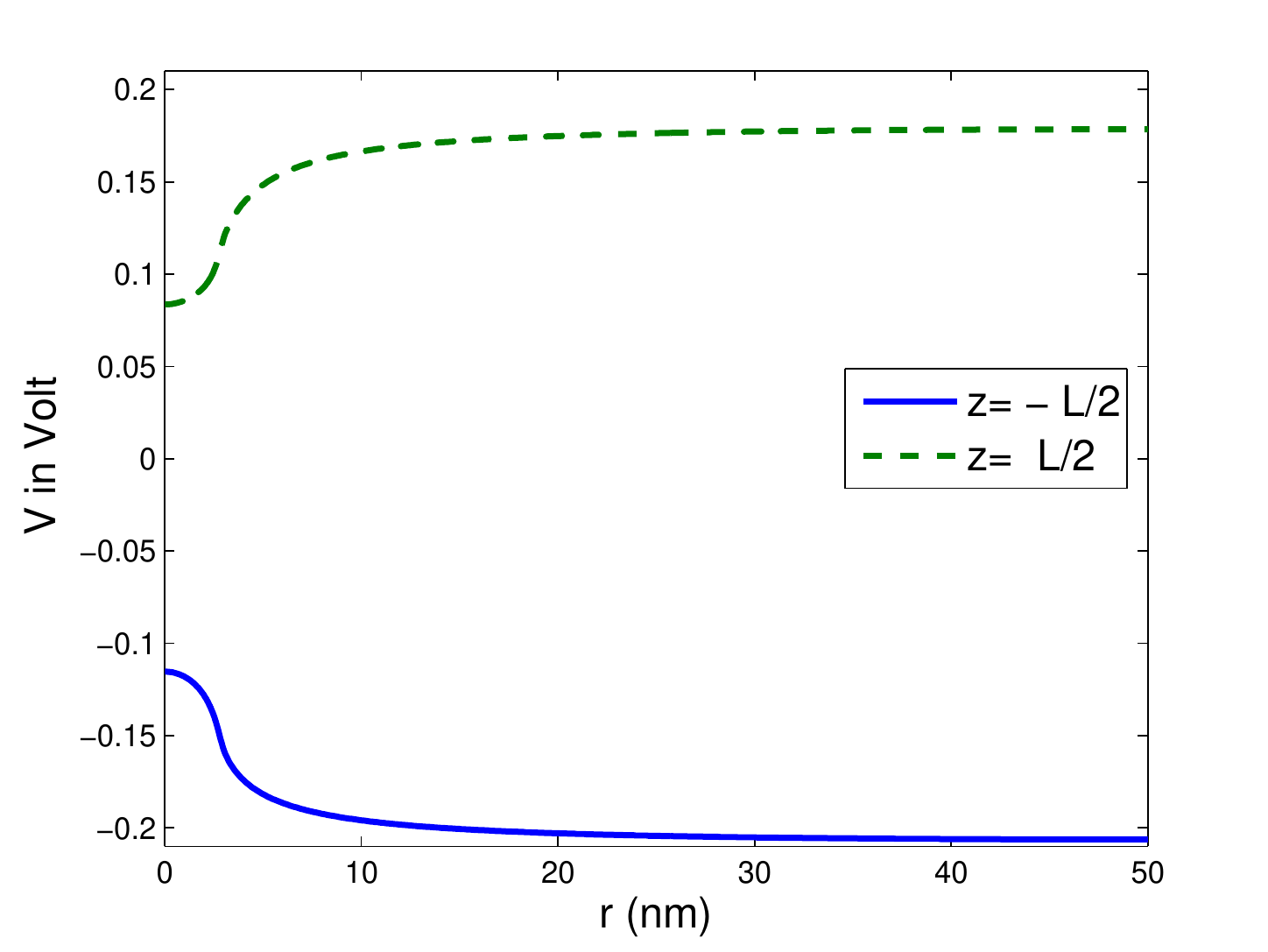}}   
\end{tabular} 
\caption{The electric potential distribution in the nanopore system. Parameters are, $\Sigma=$-0.01 C/m$^2$ $\Delta V=$0.4 V and $c_0=$0.1 M.}
\label{fig:Electric}
\end{figure}

\begin{figure}
\centering
\subfloat[Flow field in the nanopore system. The color scale indicates magnitude of velocity in $m/s$. White lines are streamlines and arrows indicate the flow direction.]{\includegraphics[width=0.70\textwidth]{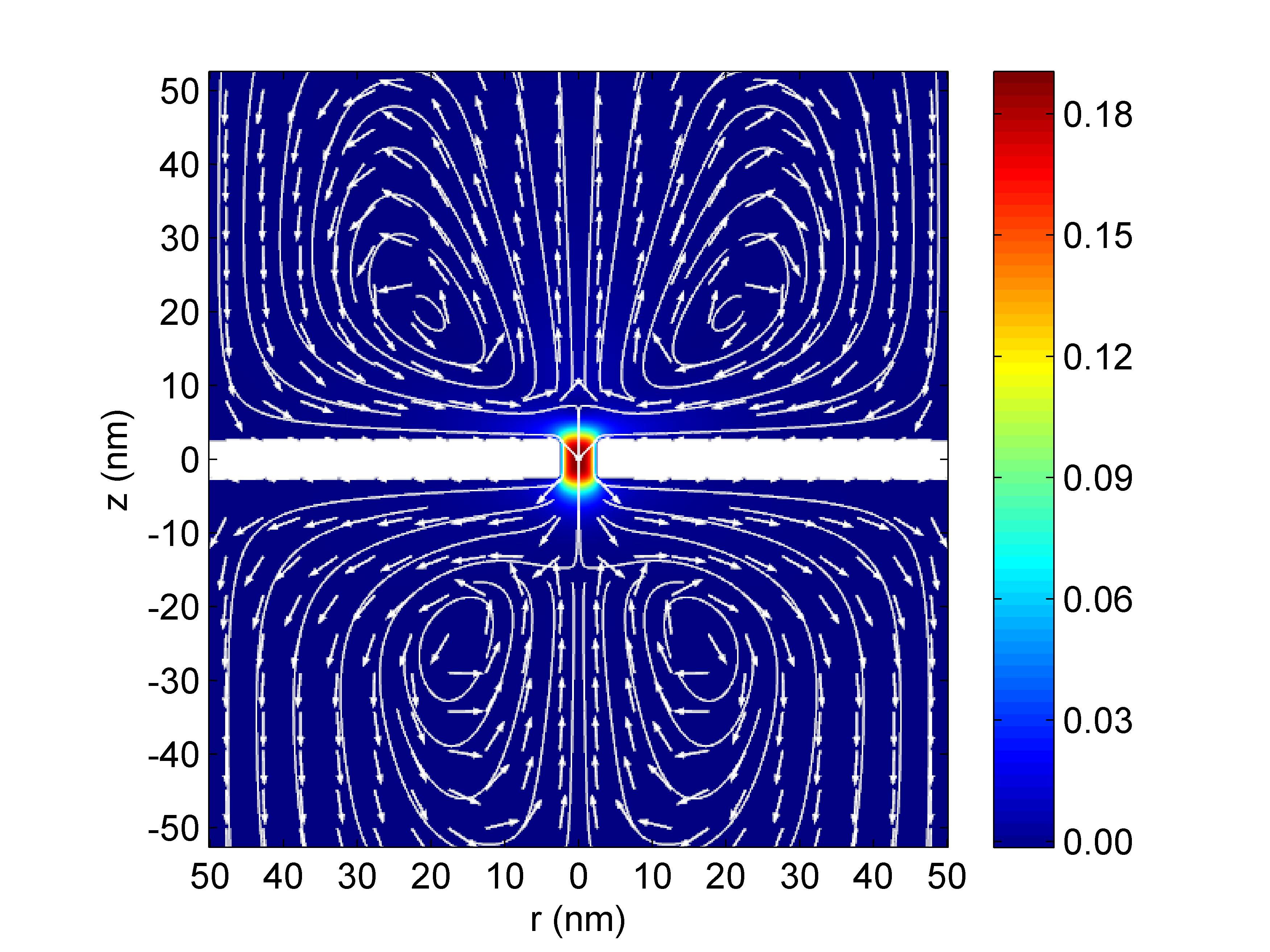}}
\\
\subfloat[Vorticity in the azimuthal direction in units of 1/s. Note that the scale is logarithmic. The vorticity due to the shear 
in the Debye layer is indicated as a flat blue color and is not represented by the color scale. The 
vorticity in this zone is in the opposite direction and is much larger than the vorticity in the bath represented by the color scale.]{\includegraphics[width=0.70\textwidth]{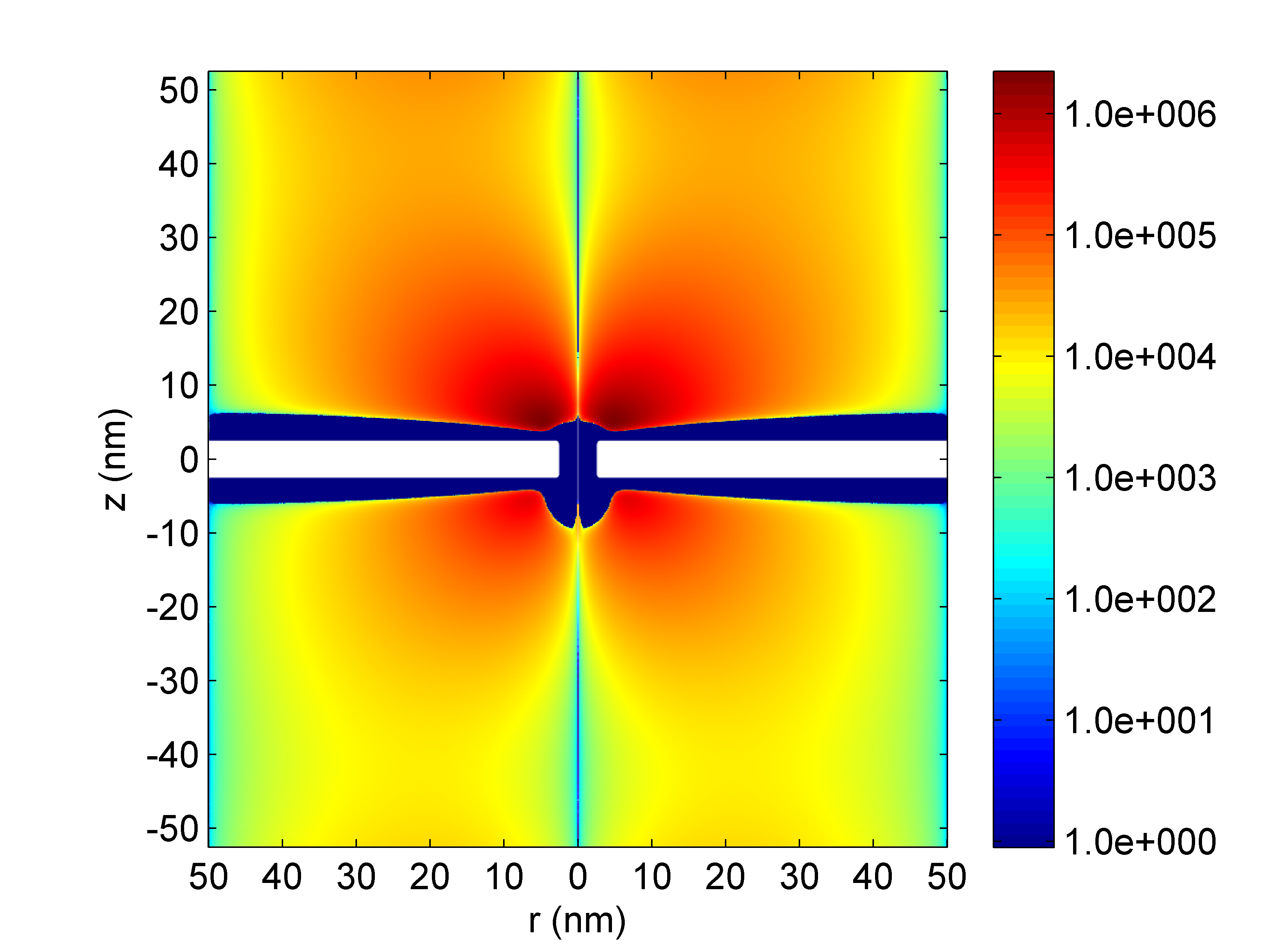}}
\caption{Hydrodynamic flow in the nanopore system. Parameters are $\Sigma=$-0.01 C/m$^2$ $\Delta V=$0.4 V and $c_0=$0.1 M.}
\label{fig:flow}
\end{figure}

The concentration distribution, electric field and flow field are shown in Figures~\ref{fig:concentration},\ref{fig:Electric} and \ref{fig:flow}
for some typical parameters. The figures correspond to a surface charge $\Sigma=$-0.01$C/m^2$, $\Delta V=$0.4 $V$ and $c_0=$0.1 $M$. The surface charge density of silica in electrolytes could vary from 0 to -0.1 $C/m^2$ under different conditions \cite{Rodrigues1999408}.

The concentration distribution (Figure \ref{fig:concentration}) for K$^+$ and Cl$^-$ are determined by both, the applied field as well as the density of fixed charges on the membrane. If the membrane had no surface charge, the K$^+$ ions would show an enhanced 
density on the $z>0$ side of the membrane and a  depletion on the $z<0$ side. The negatively charged Cl$^-$ ions would show the opposite trend, so that the two Debye layers essentially form a parallel plate capacitor with a small leakage of current through the pore. 
On the other hand, in the absence of an applied external voltage, the membrane surface charge would form a Debye layer consisting primarily of counterions. The polarity of the space charges along the membrane surfaces are determined by these two competing effects. Due to the existence of a negative surface charge on the membrane, we expect the concentration of K$^{+}$ ions on the $z>0$ 
side to be enhanced relative to that of Cl$^-$ ions. This is indeed what is observed in Figure~\ref{fig:concentration}(b), \ref{fig:concentration}(c). The relative importance of the membrane surface charge can be measured by the dimensionless parameter $\Sigma_{*} = \Sigma L / (\epsilon_s \Delta V)$ which 
represents the ratio of the intrinsic charge on the membrane to the surface charge induced by the applied field. For the situation shown in Figure~\ref{fig:concentration}, $\Sigma_{*} \approx -0.1765$. 

The potential distribution in the vicinity of the pore is shown in Figure~\ref{fig:Electric}. Clearly most of the potential drop occurs across the membrane$/$pore area, as the pore has a much larger resistance than the electrolyte solution. Thus, the electric field is the strongest in the pore region and there is a strong convergence of the electric field lines towards the pore. However, there is also a ``leakage'' of field lines into the dielectric membrane due to the fact that the conductivity of the solution is not infinite, though the field near the surface is primarily radial. The local electric field is strongly affected by the surface charges, as can be seen from the field lines near the membrane on the $z<0$ side.

The radial component of the electric field acting on the thin layer of induced charges, cause ions to move radially along the field creating electroosmotic flow. In the case depicted in Figure~\ref{fig:flow}, this flow is directed inwards (towards the pore) on the $z>0$ side of the membrane and outward (away from the pore) on the $z<0$ side. Due to the large hydrodynamic resistance the pore cannot accommodate this large radial influx and thus, mass conservation drives an axial jet away from the pore as seen in Figure~\ref{fig:flow}. The combination of the radial flow along the 
surface with the axial jet produces a toroidal vortex near the periphery of the pore. The existence of such a `corner vortex' in the reservoir near the entrance of a micro/nano channel was discussed by Yossifon {\it et al.} \cite{Yossifon2008,Yossifon2010}. We also reported these vortices in our study of the graphene nanopore using MD simulations \cite{Hu2012}. The flow vorticity which is entirely in the azimuthal direction is depicted in Figure \ref{fig:flow}(b). The vorticity in the Debye layer adjacent to the membrane is many orders of magnitude stronger than the toroidal vortex induced in the reservoir and is in the opposite direction. This Debye layer vorticity is not shown in Figure \ref{fig:flow}(b) for clarity. Also note that the vorticity scale is logarithmic.

\subsection{Electrical characteristics of the nanopore}
\begin{figure}[h]
        \centering
%
        \includegraphics[scale=0.6]{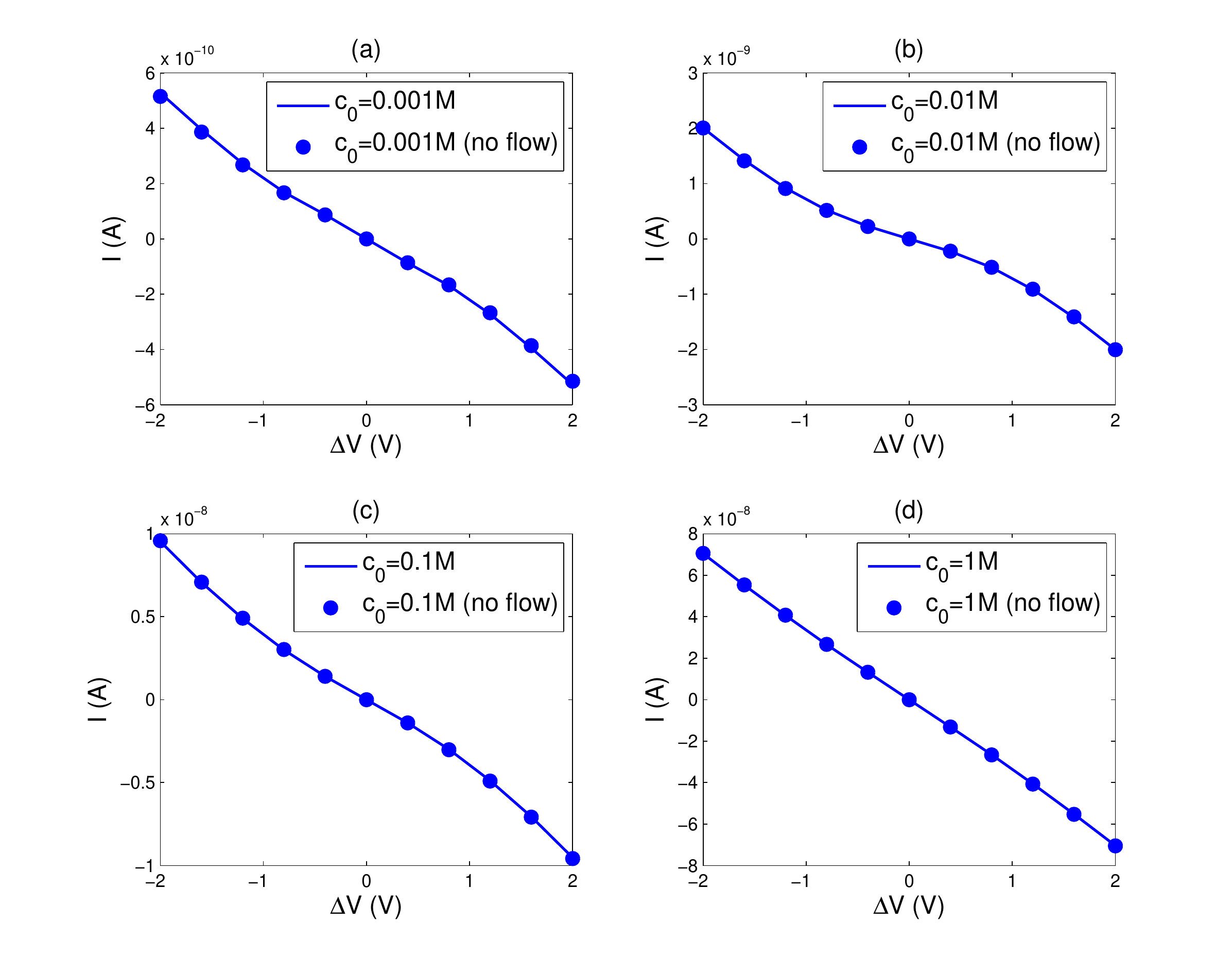}
        \caption{Current--voltage characteristics of the nanopore system for different bulk concentrations: (a) $c_0=$0.001 M (b) $c_0=$0.01 M (c) $c_0=$0.1 M (d) $c_0=$ 1.0 M. The membrane charge is fixed at $\Sigma=$-0.01 $C/m^2$.}
        \label{fig:IVall}
\end{figure}

The passage of ions through the nanopore, generates a detectable electric current. The current--voltage relation of the 
nanopore at different bulk concentrations $c_0$ (and thus different Debye lengths) is shown in Figure~\ref{fig:IVall}. 
The surface charge density is held fixed at $\Sigma=$-0.01 $C/m^2$.

From Figure \ref{fig:IVall}(d) it is seen that the nanopore has an Ohmic behavior when the bulk ion concentration is high, $c_0=1$ M. This is because at high bulk concentrations the Debye length is thin and the system can be regarded essentially as an electroneutral uniform Ohmic conductor. In this limit, the system may be regarded as a series connection of three separate resistors representing (i) the bulk resistance of the bath (ii) the access resistance of the pore and (iii) the pore resistance \cite{Kowalczyk2010,Smeets2005}: 
$R_{total}=R_{bulk}+R_{access}+R_{pore}$.
The bulk resistance is $R_{bulk}=2\rho_b L_R / (\pi L_R^2) = (2\rho_b)/(\pi L_R)$ and the access resistance is \cite{Hall1975} $R_{access}/2 = \rho_b/ (4R)$. Here $\rho_b$ is the resistivity of the solution, which is related to the mobilities of ions, 
as $\rho_b^{-1} =c_0 e^2 (D_1+D_2)/(k_B T)$.

The pore resistance is the electrical resistance ($R_c$) of the electrolyte within the electro-neutral bulk of the cylindrical pore in parallel
with a second resistance ($R_s$) due to the surface conductance of the polarized Debye layer.
Thus,  $R_{pore}=R_{c} R_{s} / ( R_{c}+R_{s} )$ where $R_{c} = (\rho_b L)/(\pi R^2)$. To calculate the surface resistance we assume that the polarized region adjacent to the surface contains only counter--ions and the pore as a whole is electroneutral. 
Then, if the electrophoretic mobility of K$^+$ is $\mu_K$, the surface resistance can be calculated as: 
$R_{s} = {L}/(2\pi R \mu_k |\Sigma |)$. Substituting the relevant numerical values, the resistance of the nanopore system when $c_0=$1 $M$ is found to be $R_{total}$ = 3.09$\times$10$^7$ $\Omega$. The resistance calculated directly from the slope of the I-V curve in Figure \ref{fig:IVall}(d)  is $R_{total}=$2.89 $\times$10$^7$ $\Omega$, in accord with the simple lumped parameter model described above. Recently, Garaj et al \cite{Garaj2010} provided conductance measurements on nanopores with diameters ranging from 5 to 23 nm in a graphene sheet. They reported good agreement with calculations based on a model that regarded the electrolyte as a homogeneous conductor. Our finding that the pore conductance at 1 molar salt may be calculated from the Ohmic model is in accord with their observations.  

At lower ionic concentrations the I--V curves in Figure \ref{fig:IVall}(a)-(c) show nonlinear behavior. However, the nonlinearity is weak and does not show the classic behavior of a limiting current leading to an over limiting current with increasing voltages as found in perm selective membranes. This may seem surprising at first because a system of densely packed nanopores is indeed expected to behave like a perm selective membrane. 
The difference can be attributed to the radial inflow of ions that prevents the formation of a fully depleted CPL next to the membrane \cite{Yossifon2010}.

\subsection{Hydrodynamic characteristics of the nanopore}
\begin{figure}
\centering
\subfloat[Total flow rate through nanopore when $\Sigma=$-0.01 $C/m^2$]{          \includegraphics[width=0.45\textwidth]{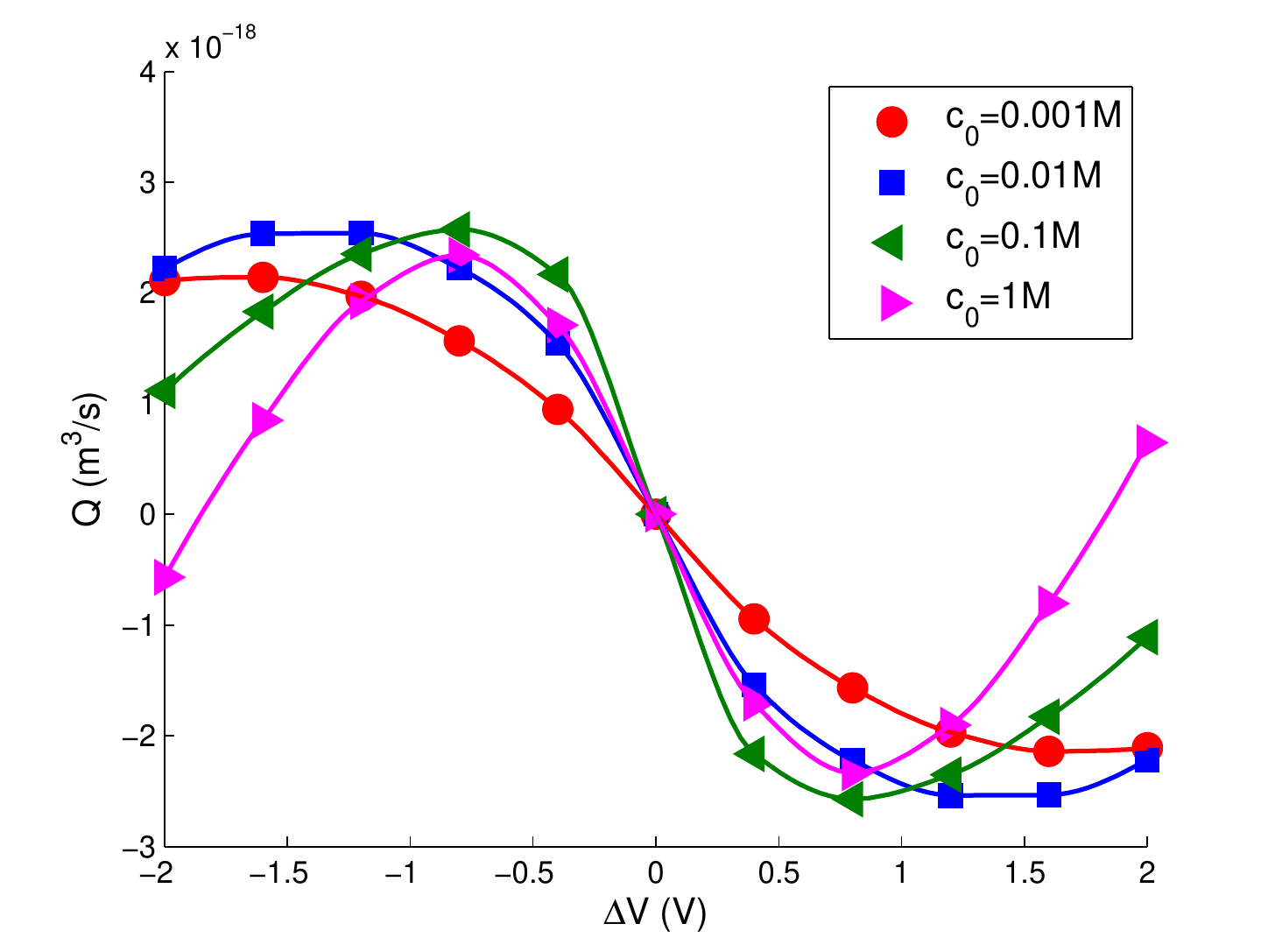}}\quad
\subfloat[Total flow rate through nanopore when $c_0=$0.1 $M$]{               \includegraphics[width=0.45\textwidth]{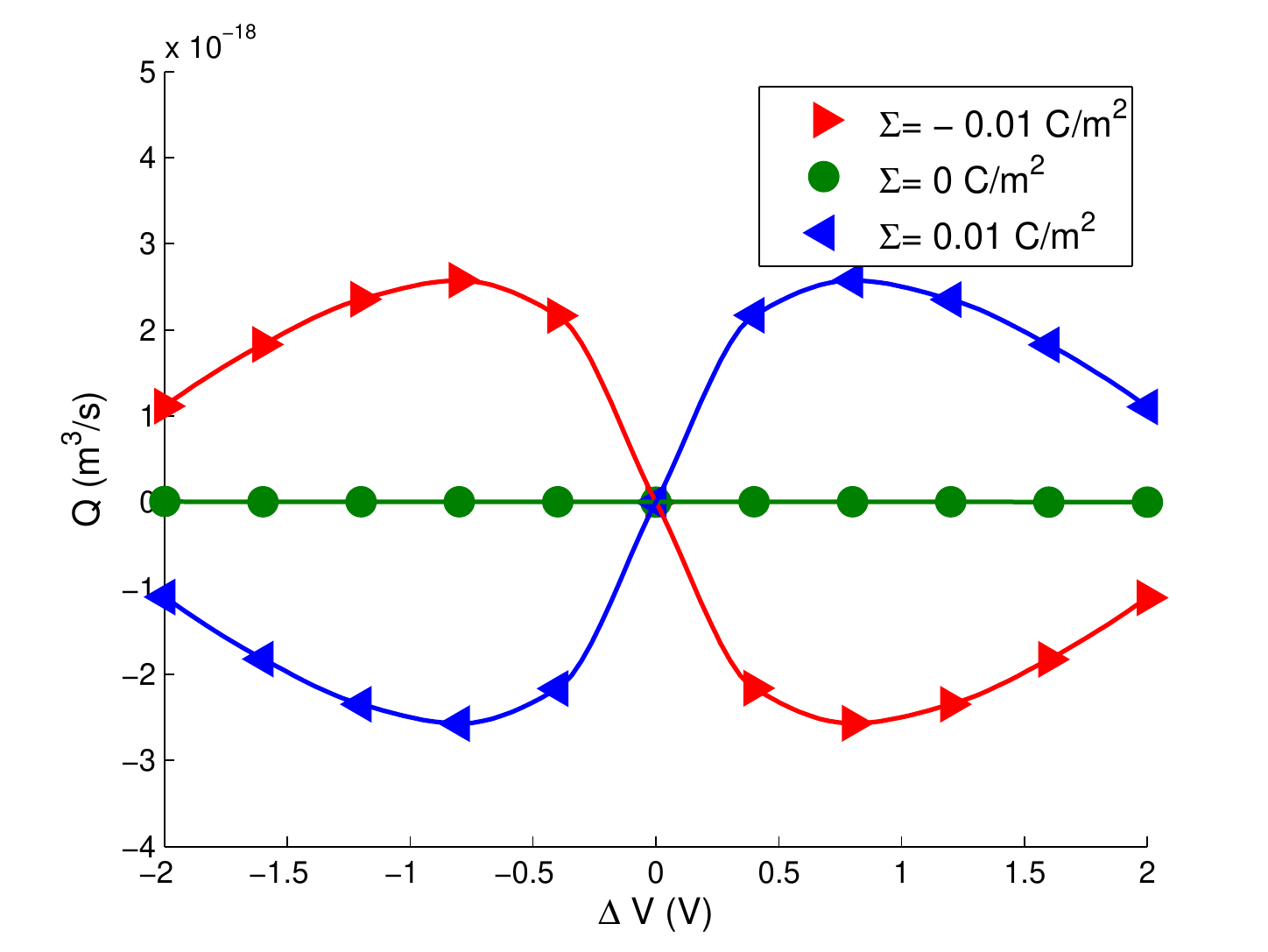}}
\caption{Flow rate through the nanopore at fixed surface charge (left) and fixed salt concentration (right).}
        \label{fig:QV}
\end{figure}

The hydrodynamic flow is driven by the electrical forces acting on the space charge adjacent to the membrane. The space charge is induced by the fixed charge on the membrane as well as by the normal component of the applied electric field. Electroosmotic flow due to the former is expected to be a linear function of the applied voltage, however, the latter generates an  `electroosmotic flow of the second kind'  in the terminology of Dukhin \cite{Dukhin1991} and has a nonlinear dependence on the voltage. As shown in Figure~\ref{fig:flow}, the fluid is driven radially along the membrane surface. Only a fraction of this radial flux crosses the membrane, the remainder being deflected into an axial jet perpendicular to the membrane. In the situation where the membrane has fixed charges, its Debye layer is enhanced on one side of the membrane and depleted on the other side by the space charge layer created by the applied field. Depending on the polarity and strength of the applied field, even the sign of the charge in the membrane's intrinsic Debye layer could reverse (refer to supplementary material). 
Thus, the strengths of the axial jets on the two sides of the membrane would generally be different for a charged membrane and there could be a net fluid flux through the pore. Such a net flow however should not appear for a membrane with no surface charge. The fluid flux through the pore was calculated by integrating the velocity over the pore area and is shown as a function of the applied voltage 
in Figure \ref{fig:QV}. As expected, (a) the flux is always zero for an uncharged membrane (b) the flux is a nonlinear function of the applied voltage but may be regarded as approximately linear for weak applied voltages (c) the flux is an odd function of the applied voltage. The last observation may be explained very simply by the fact that the transformation $\Delta V \rightarrow - \Delta V$ 
is equivalent simply to a reversal of the $z$-axis and leaves the physical system unaffected. Figure~\ref{fig:QV} also shows that the $Q-V$ characteristics are only weakly affected by the electrolyte concentration in the bath ($c_0$) which sets the Debye length $\lambda_D$. This is because the vortex in the reservoir is determined by the strength of the Helmholtz-Smoluchowski slip velocity at the edge of the Debye layer and the slip velocity itself is independent of the Debye length. In the context of Figure \ref{fig:flow}(b), changes in the Debye layer thickness would change the width of the blue zone but leaves the toroidal vortex largely unaffected. 

\subsection{Hydrodynamics induced ion selectivity}

Many biological nanopores display ion selectivity beyond what might be expected solely based on pore size \cite{Hille}. The presence of a net hydrodynamic flow through a nanopore in a charged membrane could result in an ion selectivity due to the competing effects of electrophoresis and the convective flux due to the flow. This is because only the former is sensitive to the ionic charge. 

\begin{figure}
\centering
\includegraphics[width=0.75\textwidth]{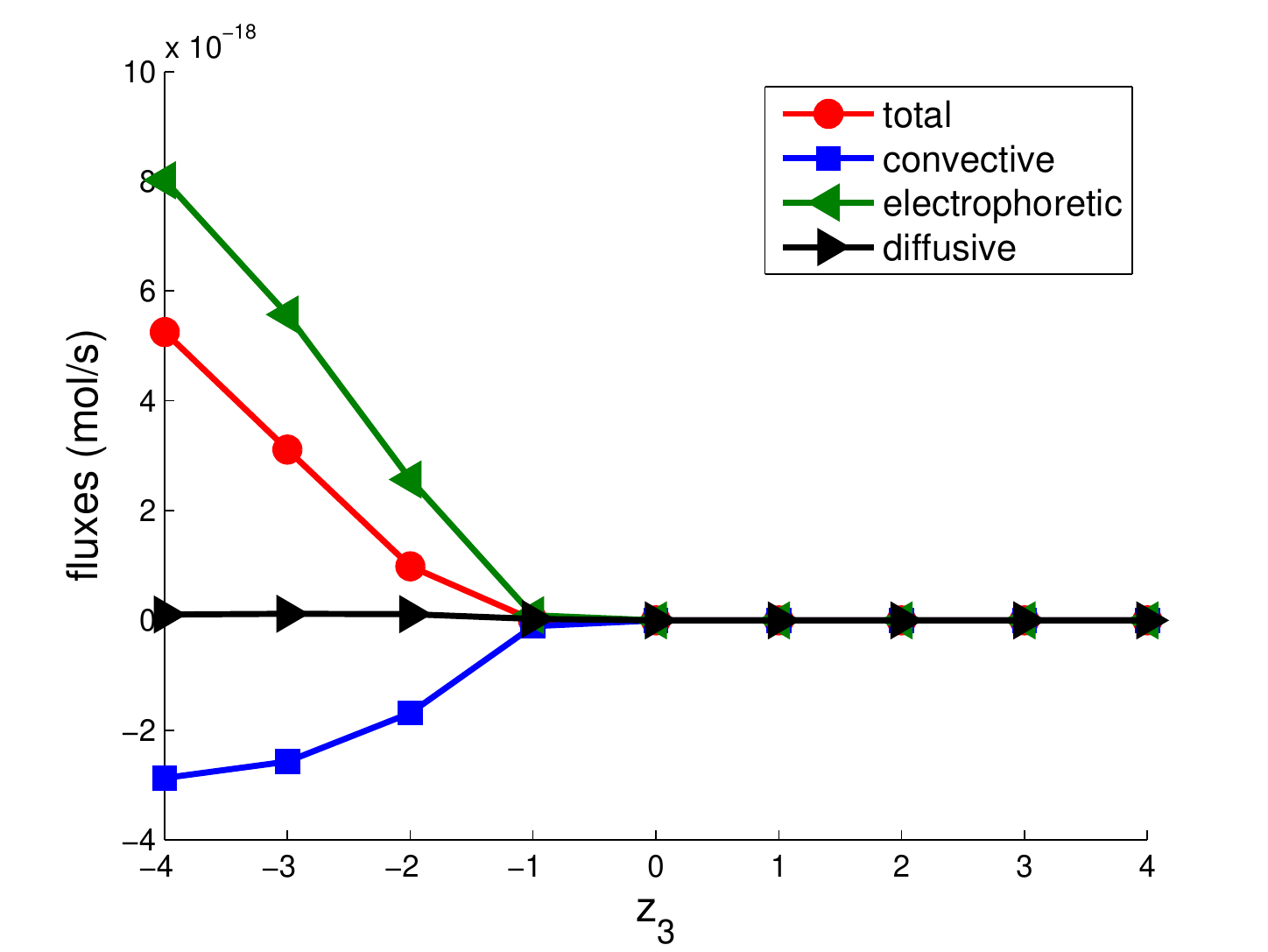}
\caption{The dependence of the various flux contributions to the total flux for a charged trace species as a function 
of its valence ($z_3$). When $z_3 <0$, the convective and electrophoretic components of the flux are comparable but in opposite 
directions. Here $c_0=$0.1$M$, $\Sigma=$-0.01 $C/m^2$ and $\Delta V=$0.5 $V$.}
\label{fig:transport}
\end{figure}

To demonstrate this effect we introduced a third charged species (concentration $c_3$) of a different valence $z_3$ into the bottom reservoir. The diffusivity of the added species is taken as $D_3=$ 4.6 $\times$10$^{-11}$ $m^2/s$, which is in the typical range of values seen in biological macromolecules and polymers. The concentration of $c_3$ is held fixed at $c_3=0.001$ M at the lower boundary and is set to zero at the upper boundary. Since this concentration is very low compared to the bulk concentration of the carrier electrolytes ($0.1$ M), it can be considered a trace species and its effect on the electric field and hydrodynamic flow neglected. Thus, we evolved $c_3$ as a passive scalar using the previously computed electric field and fluid velocity. The remaining parameters in the simulation are $\Delta V=0.5$ V and $\Sigma=-0.01$ C/m$^2$. The flux of the added species through the pore is shown in Figure \ref{fig:transport} as a function of the ion 
valence $z_3$. The flux is of course zero if $z_3 > 0$ as cations are driven away from the pore and diffusive flux is negligible. 
However, for anions, due to the competing effects of electrophoresis and convection, ions with valence in the range $0 > z_3 > -1$ are also excluded.  If an array of nanopores is fabricated to bridge a pair of parallel microfluidic channels, the pores will behave as a permselective membrane blocking all ions for which $z_3 > -1$. Such a system could therefore be used to separate biomolecules, for example proteins, using essentially the same principle as is used in the desalination of water \cite{RubinsteinDesalination}.  
Furthermore, the permselective properties can be modulated by changing the voltage applied between the microchannels. 
A similar effect was demonstrated by Karnik {\it et al} \cite{Karnik2005} to make a gate controlled `nanofluidic transistor'
though they attributed the selective properties to the Donnan potential in the channel.

\section{Discussion}

In this paper, we present a continuum level description of a problem that we had studied previously using MD simulation \cite{Hu2012}, namely, hydrodynamics and ion transport through a nanopore in a membrane. We find qualitative agreement of the two simulations: similar concentration polarization, corner vortices, nonlinear current--voltage 
characteristics as well as net flow through the nanopore. However, in the present paper, it is the membrane surface charge that creates the asymmetry responsible for the net flow. In our MD simulation, the membrane had no charge, however, the asymmetric space charge was due to a difference in mobilities between sodium and chloride ions. Here we use potassium and chlorine ions, that have almost identical mobilities. Thus, a net flow can be generated by any condition that induces an asymmetry between the two sides of the membrane. 

There are however some effects that are only captured in the MD simulation but not in our continuum simulation. One such effect is the clearly observable increase in density of water due to preferential alignment of the water molecules in the large electric field close to the pore. Since the continuum simulations are based on the incompressible fluid equations, this effect is a priori excluded. The ionic distributions in the MD simulations show a peak  a few ionic radii from the surface. This is due to a steric effect: ions cannot get closer than an ionic radius of the surface. Clearly this effect is not observable in the continuum version. 

The transport of ionic species is governed by the Nernst--Planck equation. The total flux consists of the diffusive flux, electrophoretic flux and electroosmotic flux. The relative importance of electrophoresis and electroosmosis can be determined by comparing the electrophoretic velocity and electroosmotic velocity. Both velocities reach their peak value within the pore region where the electric field is largest. In the present study the peak electrophoretic velocity is $v_{ep} \approx$ 7.7 $m/s$. From Figure \ref{fig:flow}, the maximum electroosmotic velocity is only about 0.2 $m/s$. Thus, transport of ions in the pore is dominated by electrophoresis not electroosmosis as argued by Chang {\it et al.} \cite{Chang2011}. The vortical flow could however affect the current-voltage characteristics indirectly \cite{Yossifon2010,Chang2011, ZALTZMAN2007} by modifying the ionic distributions at the pore entrance. In the current study, this effect is not of significant magnitude. The vortical flow induced in the bath could nevertheless have other significant effects. It could compete with electrophoretic transport of low mobility species such as macromolecules, as seen in Figure~\ref{fig:transport}. Another possibly significant effect that we did not study in this paper is the translocation of polyelectrolytes across pores, a very important biological process \cite{Shariff20043647} 
 that has also received much attention as a biotechnology tool \cite{KeyserOrigamiNanopore}. 
 Even though the translocation time of the polymer is set by the electric field within the pore and the pore geometry \cite{ghosal2006electrophoresis,ghosal2007effect,Ghosal2007} the rate of capture of polymers by the pore can be greatly enhanced by the toroidal vortex in the bath. This is because the polymer moves towards the pore by diffusion and one end of the polymer must insert into the pore by the process of reptation before the electric field can take hold and start the translocation. Both of these processes are very slow. The vortex can convect the polymer towards the pore as well as provide the shear needed for it to undergo a coil-stretch transition thus greatly helping the process of insertion in the pore. 

In conclusion, we can say that except for certain very specific features, most properties of the nanopore system 
are adequately described by the continuum NPP-Stokes model for fluid and ion transport. 
This is encouraging, because such continuum simulations are many orders of magnitude cheaper than the 
corresponding MD simulations and therefore represent a good compromise between accuracy and computational expenses at the mesoscale.
     
\section*{Acknowledgements}
This work was supported by grant number R01HG004842 
from the National Human Genome Research Institute,
National Institutes of Health. One of us (SG) acknowledges support from the 
Leverhulme Trust (UK).
\vskip 2cm
\bibliographystyle{unsrt}
\bibliography{references}
\end{document}